\documentclass{article}

\usepackage{arxiv}

\usepackage[utf8]{inputenc} % allow utf-8 input
\usepackage[T1]{fontenc}    % use 8-bit T1 fonts
\usepackage{hyperref}       % hyperlinks
\usepackage{url}            % simple URL typesetting
\usepackage{booktabs}       % professional-quality tables
\usepackage{amsfonts}       % blackboard math symbols
\usepackage{nicefrac}       % compact symbols for 1/2, etc.
\usepackage{microtype}      % microtypography
\usepackage{lipsum}
\usepackage{graphicx}
\usepackage{amssymb}
\usepackage{amsmath}
\usepackage{matlab-prettifier}
\usepackage[style=nature]{biblatex}
\graphicspath{ {./images/} }

\title{Fundamental Bound on Epidemic Overshoot in the SIR Model}

\author{
 Maximilian M. Nguyen \\
  Lewis-Sigler Institute\\
  Princeton University\\
  Princeton, NJ 08544 \\
  \texttt{mmnguyen@princeton.edu} \\
   \And
 Ari S. Freedman \\
  Department of Ecology and Evolutionary Biology\\
  Princeton University\\
  Princeton, NJ 08544 \\
  \texttt{arisf@princeton.edu} \\
   \And
 Sinan A. Ozbay \\
  Bendheim Center for Finance\\
  Princeton University\\
  Princeton, NJ 08544 \\
  \texttt{sozbay@princeton.edu}
  \And
 Simon A. Levin \\
  Department of Ecology and Evolutionary Biology\\
  Princeton University\\
  Princeton, NJ 08544 \\
  \texttt{slevin@princeton.edu}
}

\begin{document}
\maketitle

\begin{abstract}
We derive an exact upper bound on the epidemic overshoot for the Kermack-McKendrick SIR model. This maximal overshoot value of 0.2984... occurs at $R_0^*$ = 2.151... . In considering the utility of the notion of overshoot, a rudimentary analysis of data from the first wave of the COVID-19 pandemic in Manaus, Brazil highlights the public health hazard posed by overshoot for epidemics with $R_0$ near 2. Using the general analysis framework presented within, we then consider more complex SIR models that incorporate vaccination.
\end{abstract}

\section*{Introduction}
The overshoot of an epidemic is the proportion of the population that becomes infected after the peak of the epidemic has already passed. Formally, it is given as the difference between the fraction of the population that is susceptible at the peak of infection prevalence and at the end of the epidemic. Intuitively, it is the difference between the herd immunity threshold and the total fraction of the population that gets infected \cite{handel_what_2007, cobey_modeling_2020}. As it describes the damage to the population in the declining phase of the epidemic (i.e. when the effective reproduction number is less than 1), one might be tempted to dismiss its relative importance. However, a substantial proportion of the epidemic, and thus a large number of people, may be impacted during this phase of the epidemic dynamics.

A natural question to ask then is how large can the overshoot be and how does the overshoot depend on epidemic parameters, such as transmissibility and recovery rate? Surprisingly, this question can be answered exactly. In this paper, we first derive the bound on the overshoot in the Kermack-McKendrick limit of the SIR model \cite{kermack_contribution_1927}. We then compare the predictions of this feature of the SIR model with data taken from the first wave of the COVID-19 pandemic in Manaus, Brazil (\cite{buss_three-quarters_2021}). Beyond the basic SIR model, we then see if the bound on overshoot holds if we add additional complexity, such as vaccinations.

\section*{Results}
Over the years, the Kermack-McKendrick SIR model has become largely synonymous with the following set of ordinary differential equations (ODEs) due to their simplicity and popularity:
\begin{align}
\frac{dS}{dt}&=-\beta S I \label{eqn:Sdot}\\
\frac{dI}{dt}&=\beta S I-\gamma I \label{eqn:Idot}\\
\frac{dR}{dt}&=\gamma I
\end{align}
where $S,I$, and $R$ are the fractions of population in the susceptible, infected, or recovered state respectively. As these are the only possible states within this model, the conservation equation for the whole population is given as $S+I+R=1$. It is worth noting that the original compartmental model formulated by Kermack and McKendrick in their seminal paper from a century ago \cite{kermack_contribution_1927} is actually a more general model than the ODE model that has become synonymous with their names. The original model considered both infectiousness that depended on the amount of time since becoming infected, which has been termed age of infection, and demographic effects in the form of deaths. A considerable amount has been learned and understood in the case of the more general model that considers age of infection (see \cite{brauer_kermackmckendrick_2005, breda_formulation_2012} for an introduction), which typically takes the form of a nonlinear renewal equation. While here we have chosen to focus on the simpler ODE model, under certain assumptions our result for the overshoot can be carried over to the age-of-infection model as well.

\begin{figure}
\centering
\includegraphics[scale=0.26]{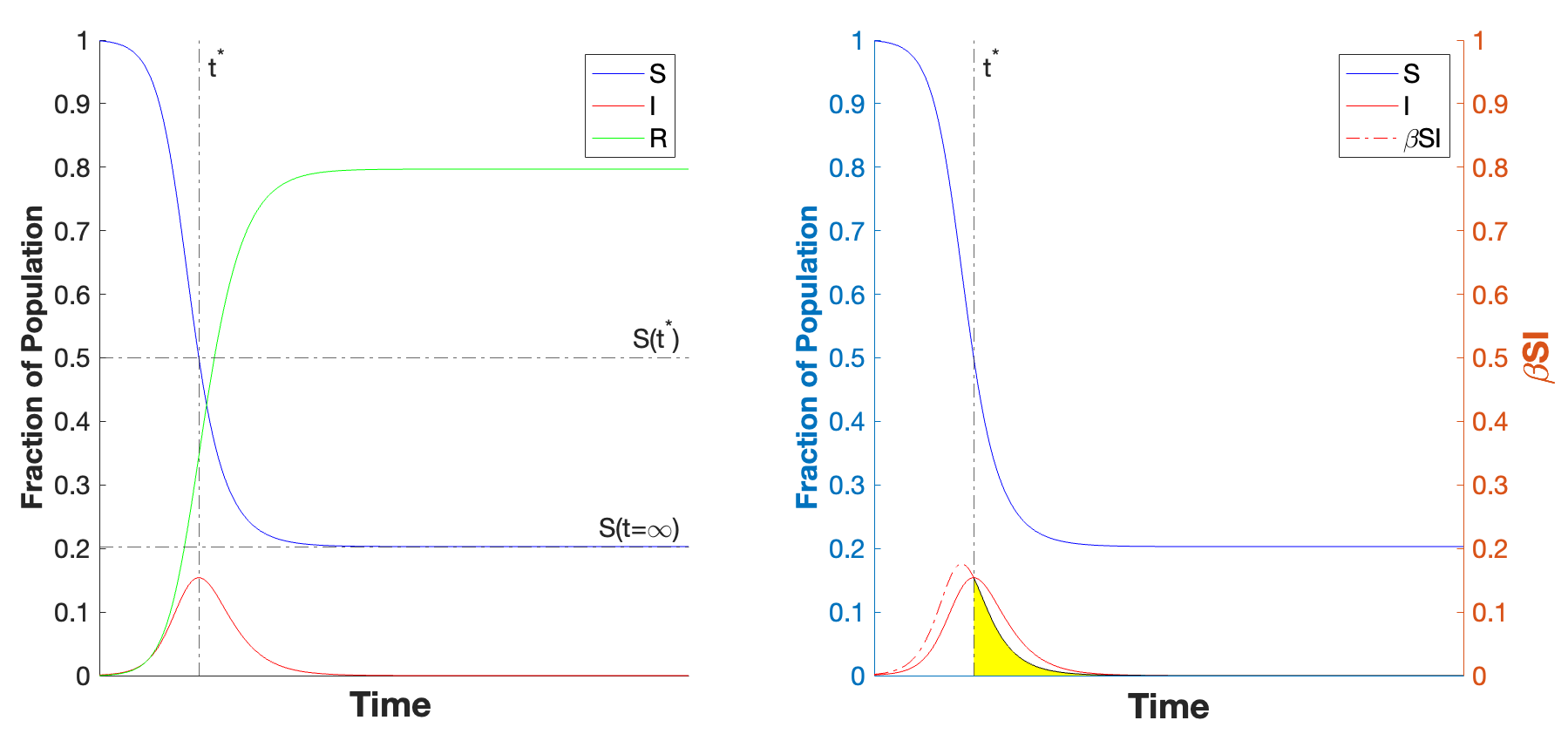}
\caption{The overshoot can be calculated in two ways: a) Overshoot is calculated as the difference between the fraction of the population that is susceptible at $t^*$ and infinite time. b) Overshoot is calculated as the integral of the infection incidence curve from $t^*$ until infinite time. Therefore overshoot corresponds to the area of the region shaded in yellow.} \label{fig:overshoot}
\end{figure}

Conceptually, the overshoot can be equivalently calculated in two ways. In the first it is given by the difference in the fraction of susceptible individuals at the peak of infection prevalence ($S_{t_*}$) and at the end of the end of the epidemic ($S_{\infty}$) (Figure \ref{fig:overshoot}a). Alternatively, it can be viewed as the integration of the number of newly infected individuals, which is given by the infection incidence rate ($\beta SI$) from the peak of infection prevalence to the end of the epidemic (Figure \ref{fig:overshoot}b). We will make use of the former relationship in the results that follow.

\begin{figure}[ht]
\centering
\includegraphics[scale=0.3]{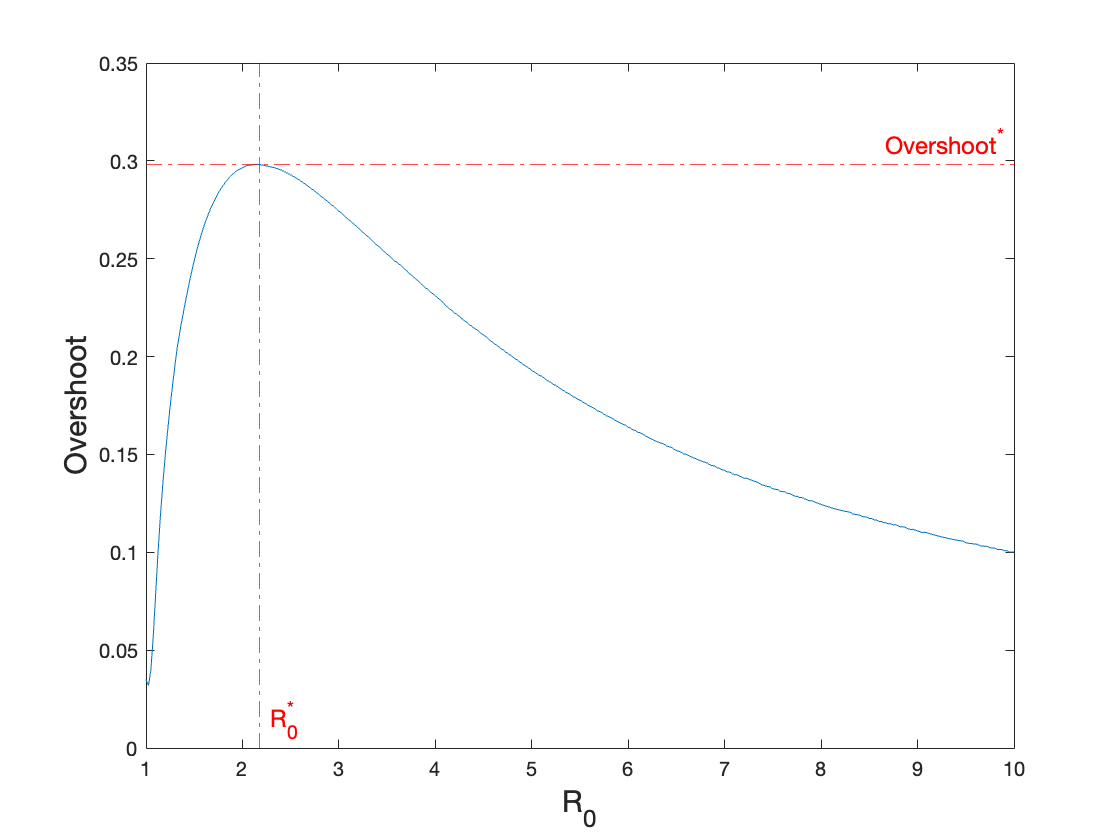}
\caption{The overshoot as a function of $R_0$ for the Kermack-McKendrick SIR model.} \label{fig:overshoot_plot}
\end{figure}

The only two parameters of the ODE model are $\beta$ and $\gamma$. A key parameter in epidemic modeling combines these two into a single parameter by taking their ratio, which is known as the basic reproduction number ($R_0$). The behavior of the overshoot can be shown to be only dependent on this single parameter, $R_0$. Plotting the dependency of overshoot on $R_0$ (Figure \ref{fig:overshoot_plot}), we observe a peak in the curve at ($R_0^*,Overshoot^*$) that sets an upper bound on the overshoot. From a public health perspective, diseases that have estimated $R_0$'s near this peak region in Figure \ref{fig:overshoot_plot} include COVID-19 (ancestral strain) \cite{billah_reproductive_2020}, SARS \cite{world_health_organization_consensus_2003}, diphtheria \cite{truelove_clinical_2020}, monkeypox \cite{grant_modelling_2020}, and ebola \cite{wong_systematic_2017}. This peak phenomena in the overshoot was first numerically observed by \cite{zarnitsyna_intermediate_2018}, though not explained. We will now derive the solution for this maximum point analytically.

\subsection*{Deriving the Exact Bound on Overshoot in the Kermack-McKendrick SIR Model}

\emph{Theorem: The maximum possible overshoot in the Kermack-McKendrick SIR model is a fraction $0.2984...$ of the entire population, with a corresponding $R_0^*=2.151...$.} \newline

\noindent Proof. \newline
Let $t^*$ be the time at the peak of the infection prevalence curve. Here we define the herd immunity threshold as the difference in the fractions of the population that are susceptible at zero time and at $t^*$. Then, the overshoot is defined as the difference in the fractions of the population that are susceptible at $t^*$ and at infinite time. This is equivalent to defining overshoot as the cumulative fraction of the population that gets infected after $t^*$.
\begin{align}
Overshoot &\equiv \int_{t^*}^{\infty} -(\frac{dS}{dt}) \,dt =\int_{t*}^{\infty} \beta S I \,dt \nonumber \\
&= S_{t*}-S_{\infty} 
\end{align}
where $S_{t*}$ and $S_{\infty}$ are the susceptible fractions at $t^*$ and infinite time respectively.
We will use $S_{t*}=\frac{1}{R_0}$ \cite{may_vaccination_1982}, which can be obtained by setting (\ref{eqn:Idot}) to zero and solving for that critical $S$. We will use the notation $X_t$ to indicate the value of compartment X at time $t$.
\begin{align}
Overshoot = \frac{1}{R_0} -S_{\infty} \label{eqn:overshoot}
\end{align} 

It is worth noting that the result that follows also holds for the more general age-of-infection model \cite{kermack_contribution_1927} if we restrict our definition of the herd immunity threshold to be the fraction of people that need to be removed from the population at the beginning of the epidemic to prevent an outbreak from occurring. While this alternative definition gives an equivalent herd immunity threshold in the ODE model where it is defined in terms of the peak of the prevalence curve, this more robust definition is needed to account for the more complicated behavior in the age-of-infection model.

Since we would like to compute maximal overshoot, we can differentiate the overshoot equation (\ref{eqn:overshoot}) with respect to $S_{\infty}$ to find the extremum. We will eliminate $R_0$ from the overshoot equation so that we have an equation only in terms of $S_{\infty}$. 

 To find an expression for $R_0$, we start by deriving the standard final size relation for the SIR model \cite{arino_final_2007,brauer_age--infection_2008}. We solve for the rate of change of I as a function of S using (\ref{eqn:Sdot})-(\ref{eqn:Idot}) to obtain
\begin{align*}
% \frac{\frac{dI}{dt}}{\frac{dS}{dt}} &= \frac{\beta SI-\gamma I}{-\beta SI} \\
{\frac{dI}{dS}} &= -1 + \frac{\gamma}{\beta S}
\end{align*}
from which it follows on integration that
$S+I - \frac{\gamma}{\beta} \ln  S$ is constant along any trajectory. 

Considering the beginning of the epidemic and the peak of the epidemic yields:
\begin{align}
S_0+I_0 - \frac{\gamma}{\beta} \ln  S_0 &= S_{\infty}+I_{\infty} - \frac{\gamma}{\beta} \ln  S_{\infty} \nonumber
\end{align}
hence
\begin{align}
\frac{\beta}{\gamma} (S_{\infty}-S_0 + I_{\infty}-I_0) &= \ln(\frac{ S_{\infty}}{S_0}) \label{eqn:finalsizepre}
\end{align}

We now define the initial conditions: $S_0=1-\epsilon$ and $I_0=\epsilon$, where $\epsilon$ is the (infinitesimally small) fraction of initially infected individuals. We assume that the number of initially infected individuals ($\epsilon$) is much smaller than the size of the population (i.e., $\epsilon << 1$). For the scale that we have in mind, such as those of city populations and larger, it is thus reasonable to make the approximation $1-\epsilon \approx 1$. We also use the standard asymptotic of the SIR model that there are no infected individuals at the end of an SIR epidemic: $I_{\infty} = 0$. Taking the above conditions together and recalling that $R_0 = \frac{\beta}{\gamma}$, we obtain that
\begin{align}
S_{\infty} &= e^{R_0 (S_{\infty}-1)} \label{eqn:s_infty_trans}
\end{align}
The resulting equation (\ref{eqn:s_infty_trans}) is the final size relation for the Kermack-McKendrick SIR model. Importantly, this final size relation taken together with the alternative definition for the herd immunity threshold implies the subsequent result for overshoot holds not only for the simpler ODE model considered here, but also for the more general age-of-infection model of Kermack and McKendrick \cite{kermack_contribution_1927}. The robustness of the final size relation in the context of the more general model can be more easily viewed through the lens of a renewal equation for the force of infection, see \cite{thieme_mathematics_2018, breda_formulation_2012, diekmann_mathematical_2013, brauer_age--infection_2008} for a derivation and a more complete discussion. 

Rearranging for $R_0$ yields the following expression:
\begin{align}
\frac{\ln (S_{\infty})}{S_{\infty}-1} &= R_0  \label{eqn:R0Expression}
\end{align}
We then substitute this $R_0$ expression (\ref{eqn:R0Expression}) into the overshoot equation (\ref{eqn:overshoot}).
\begin{align}
Overshoot = \frac{S_{\infty}-1}{\ln (S_{\infty})} -S_{\infty} \label{eqn:overshoot_sub}
\end{align} 
Differentiating with respect to $S_{\infty}$ and setting the equation to zero to find the maximum overshoot yields:
\begin{align}
(\ln S_{\infty*})^2 = \ln S_{\infty*}-1+\frac{1}{S_{\infty*}}
\end{align}
whose solution is 
\begin{align}
S_{\infty*} = 0.1664... \nonumber
\end{align}
and which corresponds to
\begin{align}
Overshoot^* &= 0.2984... \label{eqn:upperbound}
\end{align}
using (\ref{eqn:overshoot_sub}). The corresponding $R_0$ calculated using (\ref{eqn:R0Expression}) is
\begin{align}
R_0^* &= 2.151... \label{eqn:R0_answer}
\end{align} 
This concludes the proof $\blacksquare$

Additionally, to find the total recovered fraction is straightforward. In the asymptotic limit of the SIR model, there are no remaining infected individuals, so $R_{\infty*} = 1-S_{\infty*}$.
\begin{align}
R_{\infty*} = 1-0.1664... = 0.8336...
\end{align}
In other words, approximately 5 out of every 6 individuals in the population will have experience infection when overshoot is maximized.

\section*{Discussion}
We have proved that the maximum fraction of the population that can be infected during the overshoot phase of an epidemic in the Kermack-McKendrick SIR model is just under $0.3$, with a corresponding basic reproduction number of $R_0\approx2.15$. 

Given the clear predictions of this feature of the SIR model, it is reasonable to ask whether the theory matches any real-world epidemics. While high-quality data on large, unmitigated epidemics (for which the SIR model would most directly apply) in human populations is rare, we will now perform a rudimentary analysis of data from the first wave of the COVID-19 pandemic in Manaus, Brazil as given by Buss et al. (\cite{buss_three-quarters_2021}). While the city did implement some small level of non-pharmaceutical interventions, for the purpose of calculation let us take at face value that the epidemic spread through the city practically unmitigated.

To estimate the theoretical prediction of overshoot in the SIR model, we need to first estimate $R_0$. The conservative, forward-looking approach we take here is to take the maximum of the effective reproduction number ($R_t$) when the epidemic is first starting. Using data from Buss et al. for $R_t$ in Manaus as a function of date of symptom onset, which we take as a proxy for time (\cite{buss_three-quarters_2021}), the $R_0$ was approximately 2.3 in Manaus in mid-March (Figure \ref{fig:Manaus_Rt}). For $R_0=2.3$, using Figure \ref{fig:overshoot_plot} as a reference, the theoretical prediction for overshoot is approximately $29\%$. Thus, if $R_0$ can be estimated early on in the epidemic, the overshoot can be subsequently predicted within the context of an SIR model before the peak of the epidemic occurs, which in practice provides more time for public health measures and interventions to be implemented before the overshoot phase takes place.

To calculate the overshoot as observed directly from the data, we again refer to the time series data for $R_t$ (Figure \ref{fig:Manaus_Rt}). We will consider the time when $R_t=1$ to be when the epidemic peaks ($t^*$). Reading the data suggests the first COVID-19 wave peaked in late April. We note that $R_t$ stays around 1 until mid-August, when it starts rising again. As the basic SIR model does not consider such complex late-time behavior, for the purpose of this analysis, we will consider the first wave to have ended by mid-August. We note that assigning an endpoint to the data is a strong assumption, and that actually determining the turning and end point of an epidemic in the context of epidemic forecasting is not a simple matter \cite{castro_turning_2020}. 

With the date of an epidemic peak in hand, we now turn to reading the prevalence curve. Specifically, we will be using the mean data given by seroreversion-adjusted prevalence at a 1.4 S/C threshold for positive detection (Figure \ref{fig:seroprev}), which is adapted from Buss et al. \cite{buss_three-quarters_2021}. The seroreversion adjustment is their best attempt for controlling for antibody waning. Given this correction, we will take this curve as the cumulative outbreak size. The 1.4 S/C threshold is based on the sampling threshold in relative light units for deciding whether a sample has a significant positive chemiluminescence signal over the calibration. After fitting the time series points to a simple logistic curve, it can be seen that when $R_t$ first reached 1 (indicating the epidemic had peaked), the cumulative fraction of the population that had been infected was approximately $36\%$. From here, we see that the cumulative fraction that becomes infected between this time point when $R_t$ first reached 1 and the end of the first wave in mid-August (i.e. the overshoot) is $30\%$ from the data. 

We therefore see that the SIR model prediction for overshoot aligns with the value derived from the data, suggesting that the dynamics of the first wave of COVID-19 in Manaus, Brazil can be approximated by a simple SIR model. While the crude analysis above makes several strong assumptions about the nature of the unmitigated spread, the endpoint of the wave, the accuracy of the seroprevalence testing and correction methods, and the fidelity of the sampling intervals, the fit between the data and a SIR model is perhaps unsurprising given the relatively high population density of Manaus and general lack of thorough mitigation measures. To a first-order approximation, the data suggests that overshoot indeed poses a significant amount of public health hazard when the $R_0$ is in the neighborhood of 2. And that for well-mixed, unmitigated epidemics that may be approximated by SIR dynamics, overshoot may be a sizeable portion of the dynamics and overall attack rate.

The mathematical intuition on why there is a peak in the overshoot as a function of $R_0$ can be seen by inspection of Equation \ref{eqn:overshoot}. The first term, $\frac{1}{R_0}$, monotonically decreases with increasing $R_0$ . The last term, $-S_{\infty}$, monotonically increases with $R_0$. Thus a trade-off in the two terms results in a single intermediate peak. The epidemiological intuition behind a peak in the overshoot is that the total number of individuals infected during the epidemic grows monotonically with increasing $R_0$. However, too high of an $R_0$ leads to a sharp growth in the number of infected individuals, which burns through most of the population before the infection prevalence peak is reached, leaving few susceptible individuals left for the overshoot phase. This is seen by a monotonic decrease in the fraction of infected individuals that occur in the overshoot phase with increased $R_0$ (Figure \ref{fig:over_R0}). Thus the maximal overshoot occurs as a trade-off between those two directions. It is interesting to note that while the overshoot is a non-monotonic function of $R_0$, in contrast, the ratio of overshoot to outbreak size is a strictly decreasing function of $R_0$ (see Supplemental Information for further discussion).

The fundamental upper bound on the overshoot derived here also seems to hold under the addition of more complexity into the SIR model (see Supplemental Information). Upon the addition of different modes of vaccination, we find the bound on overshoot still holds in all cases considered. In the 2-strain with vaccination SIR model of Zarnitsyna et al. \cite{zarnitsyna_intermediate_2018}, the overshoot depends on both the level of strain dominance and vaccination rate, but from their results it is numerically seen that any amount of vaccination will produce an overshoot lower than the bound found here. Different control measures and strategies may reduce the overshoot as compared to the unmitigated case \cite{handel_what_2007}, keeping this upper bound intact. Future work may explore how general this bound is for SIR models with other types of complexities or for models beyond the SIR-type.

\section*{Author Contributions}
M.M.N., A.S.F., S.A.O., S.A.L. designed research, performed research, and wrote and reviewed the manuscript.

\section*{Acknowledgements}
The authors would like to acknowledge Bryan Grenfell and Chadi Saad-Roy for their useful suggestions. 

\section*{Data Accessibility}
Code to generate Results and Figures is given in the Supplemental Materials.

\section*{Funding Statement}
The authors would like to acknowledge generous funding support provided by the National Science Foundation (CCF1917819 and CNS-2041952), the Army Research Office (W911NF-18-1-0325), and a gift from the William H. Miller III 2018 Trust. The authors declare no competing interests.

\printbibliography

@article{kermack_contribution_1927,
	title = {A contribution to the mathematical theory of epidemics},
	volume = {115},
	url = {https://royalsocietypublishing.org/doi/abs/10.1098/rspa.1927.0118},
	doi = {10.1098/rspa.1927.0118},
	pages = {700--721},
	number = {772},
	journaltitle = {Proceedings of the Royal Society of London. Series A, Containing Papers of a Mathematical and Physical Character},
	author = {Kermack, William Ogilvy and {McKendrick}, A. G. and Walker, Gilbert Thomas},
	urldate = {2022-09-16},
	date = {1927-08},
	note = {Publisher: Royal Society},
}

@article{handel_what_2007,
	title = {What is the best control strategy for multiple infectious disease outbreaks?},
	volume = {274},
	url = {https://royalsocietypublishing.org/doi/10.1098/rspb.2006.0015},
	doi = {10.1098/rspb.2006.0015},
	pages = {833--837},
	number = {1611},
	journaltitle = {Proceedings of the Royal Society B: Biological Sciences},
	author = {Handel, Andreas and Longini, Ira M and Antia, Rustom},
	urldate = {2022-11-10},
	date = {2007-03-22},
	note = {Publisher: Royal Society},
}

@article{zarnitsyna_intermediate_2018,
	title = {Intermediate levels of vaccination coverage may minimize seasonal influenza outbreaks},
	volume = {13},
	issn = {1932-6203},
	url = {https://journals.plos.org/plosone/article?id=10.1371/journal.pone.0199674},
	doi = {10.1371/journal.pone.0199674},
	pages = {e0199674},
	number = {6},
	journaltitle = {{PLOS} {ONE}},
	shortjournal = {{PLOS} {ONE}},
	author = {Zarnitsyna, Veronika I. and Bulusheva, Irina and Handel, Andreas and Longini, Ira M. and Halloran, M. Elizabeth and Antia, Rustom},
	urldate = {2022-11-10},
	date = {2018-06-26},
	langid = {english},
	note = {Publisher: Public Library of Science},
}

@article{cobey_modeling_2020,
	title = {Modeling infectious disease dynamics},
	volume = {368},
	url = {https://www.science.org/doi/10.1126/science.abb5659},
	doi = {10.1126/science.abb5659},
	pages = {713--714},
	number = {6492},
	journaltitle = {Science},
	author = {Cobey, Sarah},
	urldate = {2022-11-10},
	date = {2020-05-15},
	note = {Publisher: American Association for the Advancement of Science},
}

@article{brauer_age--infection_2008,
	title = {Age-of-infection and the final size relation},
	volume = {5},
	pages = {681},
	number = {4},
	journaltitle = {Mathematical Biosciences \& Engineering},
	author = {Brauer, Fred},
	date = {2008},
	note = {Publisher: American Institute of Mathematical Sciences},
}

@article{arino_final_2007,
	title = {A final size relation for epidemic models},
	volume = {4},
	pages = {159},
	number = {2},
	journaltitle = {Mathematical Biosciences \& Engineering},
	author = {Arino, Julien and Brauer, Fred and van den Driessche, Pauline and Watmough, James and Wu, Jianhong},
	date = {2007},
	note = {Publisher: American Institute of Mathematical Sciences},
}

@article{may_vaccination_1982,
	title = {Vaccination programmes and herd immunity},
	volume = {300},
	rights = {1982 Nature Publishing Group},
	issn = {1476-4687},
	url = {https://www.nature.com/articles/300481a0},
	doi = {10.1038/300481a0},
	pages = {481--483},
	number = {5892},
	journaltitle = {Nature},
	author = {May, Robert M.},
	urldate = {2022-11-10},
	date = {1982-12},
	langid = {english},
	note = {Number: 5892
Publisher: Nature Publishing Group},
}

@article{grant_modelling_2020,
	title = {Modelling human-to-human transmission of monkeypox},
	volume = {98},
	issn = {0042-9686},
	url = {https://www.ncbi.nlm.nih.gov/pmc/articles/PMC7463189/},
	doi = {10.2471/BLT.19.242347},
	pages = {638--640},
	number = {9},
	journaltitle = {Bulletin of the World Health Organization},
	shortjournal = {Bull World Health Organ},
	author = {Grant, Rebecca and Nguyen, Liem-Binh Luong and Breban, Romulus},
	urldate = {2022-12-08},
	date = {2020-09-01},
	pmid = {33012864},
	pmcid = {PMC7463189},
}

@article{billah_reproductive_2020,
	title = {Reproductive number of coronavirus: A systematic review and meta-analysis based on global level evidence},
	volume = {15},
	issn = {1932-6203},
	url = {https://journals.plos.org/plosone/article?id=10.1371/journal.pone.0242128},
	doi = {10.1371/journal.pone.0242128},
	shorttitle = {Reproductive number of coronavirus},
	pages = {e0242128},
	number = {11},
	journaltitle = {{PLOS} {ONE}},
	shortjournal = {{PLOS} {ONE}},
	author = {Billah, Md Arif and Miah, Md Mamun and Khan, Md Nuruzzaman},
	urldate = {2022-12-08},
	date = {2020-11-11},
	langid = {english},
	note = {Publisher: Public Library of Science},
}

@report{world_health_organization_consensus_2003,
	title = {Consensus document on the epidemiology of severe acute respiratory syndrome ({SARS})},
	url = {https://apps.who.int/iris/handle/10665/70863},
	number = {{WHO}/{CDS}/{CSR}/{GAR}/2003.11},
	institution = {World Health Organization},
	author = {{World Health Organization}},
	urldate = {2022-12-08},
	date = {2003},
	langid = {english},
	note = {number-of-pages: 46},
}

@article{wong_systematic_2017,
	title = {A systematic review of early modelling studies of Ebola virus disease in West Africa},
	volume = {145},
	issn = {0950-2688, 1469-4409},
	url = {https://www.cambridge.org/core/journals/epidemiology-and-infection/article/systematic-review-of-early-modelling-studies-of-ebola-virus-disease-in-west-africa/154353B9A815326FE3656046AD6390B6},
	doi = {10.1017/S0950268817000164},
	pages = {1069--1094},
	number = {6},
	journaltitle = {Epidemiology \& Infection},
	author = {Wong, Z. S. Y. and Bui, C. M. and Chughtai, A. A. and Macintyre, C. R.},
	urldate = {2022-12-08},
	date = {2017-04},
	langid = {english},
	note = {Publisher: Cambridge University Press},
}

@article{truelove_clinical_2020,
	title = {Clinical and Epidemiological Aspects of Diphtheria: A Systematic Review and Pooled Analysis},
	volume = {71},
	issn = {1058-4838},
	url = {https://doi.org/10.1093/cid/ciz808},
	doi = {10.1093/cid/ciz808},
	shorttitle = {Clinical and Epidemiological Aspects of Diphtheria},
	pages = {89--97},
	number = {1},
	journaltitle = {Clinical Infectious Diseases},
	shortjournal = {Clinical Infectious Diseases},
	author = {Truelove, Shaun A and Keegan, Lindsay T and Moss, William J and Chaisson, Lelia H and Macher, Emilie and Azman, Andrew S and Lessler, Justin},
	urldate = {2022-12-08},
	date = {2020-06-24},
}

@book{thieme_mathematics_2018,
	title = {Mathematics in Population Biology},
	isbn = {978-0-691-18765-5},
	pagetotal = {564},
	publisher = {Princeton University Press},
	author = {Thieme, Horst R.},
	date = {2018-06-05},
	langid = {english},
	note = {Google-Books-{ID}: 9f9aDwAAQBAJ},
}

@article{breda_formulation_2012,
	title = {On the formulation of epidemic models (an appraisal of Kermack and {McKendrick})},
	volume = {6},
	issn = {1751-3758},
	url = {https://doi.org/10.1080/17513758.2012.716454},
	doi = {10.1080/17513758.2012.716454},
	pages = {103--117},
	issue = {sup2},
	journaltitle = {Journal of Biological Dynamics},
	author = {Breda, D. and Diekmann, O. and de Graaf, W.   F. and Pugliese, A. and Vermiglio, R.},
	urldate = {2023-08-18},
	date = {2012-09-01},
	pmid = {22897721},
	note = {Publisher: Taylor \& Francis
\_eprint: https://doi.org/10.1080/17513758.2012.716454},
}

@book{diekmann_mathematical_2013,
	title = {Mathematical Tools for Understanding Infectious Disease Dynamics},
	isbn = {978-0-691-15539-5},
	pagetotal = {516},
	publisher = {Princeton University Press},
	author = {Diekmann, Odo and Heesterbeek, Hans and Britton, Tom},
	date = {2013},
	langid = {english},
}

@article{buss_three-quarters_2021,
	title = {Three-quarters attack rate of {SARS}-{CoV}-2 in the Brazilian Amazon during a largely unmitigated epidemic},
	volume = {371},
	url = {https://www.science.org/doi/full/10.1126/science.abe9728},
	doi = {10.1126/science.abe9728},
	pages = {288--292},
	number = {6526},
	journaltitle = {Science},
	author = {Buss, Lewis F. and Prete, Carlos A. and Abrahim, Claudia M. M. and Mendrone, Alfredo and Salomon, Tassila and de Almeida-Neto, Cesar and França, Rafael F. O. and Belotti, Maria C. and Carvalho, Maria P. S. S. and Costa, Allyson G. and Crispim, Myuki A. E. and Ferreira, Suzete C. and Fraiji, Nelson A. and Gurzenda, Susie and Whittaker, Charles and Kamaura, Leonardo T. and Takecian, Pedro L. and da Silva Peixoto, Pedro and Oikawa, Marcio K. and Nishiya, Anna S. and Rocha, Vanderson and Salles, Nanci A. and de Souza Santos, Andreza Aruska and da Silva, Martirene A. and Custer, Brian and Parag, Kris V. and Barral-Netto, Manoel and Kraemer, Moritz U. G. and Pereira, Rafael H. M. and Pybus, Oliver G. and Busch, Michael P. and Castro, Márcia C. and Dye, Christopher and Nascimento, Vítor H. and Faria, Nuno R. and Sabino, Ester C.},
	urldate = {2023-08-18},
	date = {2021-01-15},
	note = {Publisher: American Association for the Advancement of Science},
}

@article{brauer_kermackmckendrick_2005,
	title = {The Kermack–{McKendrick} epidemic model revisited},
	volume = {198},
	issn = {0025-5564},
	url = {https://www.sciencedirect.com/science/article/pii/S0025556405001331},
	doi = {10.1016/j.mbs.2005.07.006},
	pages = {119--131},
	number = {2},
	journaltitle = {Mathematical Biosciences},
	shortjournal = {Mathematical Biosciences},
	author = {Brauer, Fred},
	urldate = {2023-10-11},
	date = {2005-12-01},
}

@article{castro_turning_2020,
	title = {The turning point and end of an expanding epidemic cannot be precisely forecast},
	volume = {117},
	url = {https://www.pnas.org/doi/abs/10.1073/pnas.2007868117},
	doi = {10.1073/pnas.2007868117},
	pages = {26190--26196},
	number = {42},
	journaltitle = {Proceedings of the National Academy of Sciences},
	author = {Castro, Mario and Ares, Saúl and Cuesta, José A. and Manrubia, Susanna},
	urldate = {2023-10-16},
	date = {2020-10-20},
	note = {Publisher: Proceedings of the National Academy of Sciences},
}

\newpage
\begin{center}
\textbf{\large Supplemental Materials: Fundamental Bound on Epidemic Overshoot in the SIR Model}
\end{center}

\subsection*{Upper Bounds on Overshoot in Models that Include Vaccinations}
Beyond the Kermack-McKendrick SIR model, one can ask if the bound on overshoot still holds if other complexities are added to the model. First, we will consider the addition of vaccinations. 

We will consider three qualitatively different types of curves for the vaccination rate (Figure \ref{fig:dVdt}). These correspond to different scenarios that might be modeled. The first model assumes a vaccination rate of zero after the outbreak begins, which implies all vaccinations occur before the outbreak. The second model of vaccination assumes a constant per-capita vaccination rate. This is a situation where all susceptible individuals get vaccinated at the same rate. This assumption yields a vaccination curve for the population that is concave down. The third type of model assumes a risk-driven vaccination rate that depends on the number of infected individuals. This yields a non-monotonic vaccination curve for the population that switches from being initially concave up to being concave down. Depending on the scenario being analyzed, one model might be more appropriate to use than others. Below we discuss each model in further detail by providing the corresponding system of equations, relevant scenarios the model might correspond to in reality, and the corresponding maximal overshoot for each model.

\noindent \textit{Maximal Overshoot when the Number of Vaccinated Individuals is Constant}

The first model of vaccination assumes there are no vaccinations during the outbreak, which implies a fixed number of vaccinated individuals over the course of the epidemic. Such a scenario might be the reintroduction of an infectious disease into a population that has a pre-existing level of immunity. 

Since the number of vaccinated individuals is constant, this implies all vaccinations occurred prior to the initial time step. The calculation is then trivial assuming vaccinations provide complete and permanent immunity. In that case, vaccinated individuals can simply be ignored entirely in the dynamics, resulting in the maximal overshoot simply scaling with the unvaccinated fraction.
\begin{align}
Overshoot^*_{SIRV}=(1-V)0.2984...
\end{align}

\noindent \textit{Maximal Overshoot Under Addition of Constant Per-Capita Vaccination}

We next consider a more typical scenario where the vaccination rate per unvaccinated individual is constant per unit time. Barring any additional information about the population or the epidemic, it is reasonable to assume that all susceptible individuals are vaccinated at the same rate. Consider the following SIRV model: 
\begin{align}
\frac{dS}{dt}&=-\beta S I -\lambda S \label{eqn:SdotSIRV1}\\
\frac{dI}{dt}&=\beta S I-\gamma I \\
\frac{dR}{dt}&=\gamma I \\
\frac{dV}{dt} &= \lambda S 
\end{align}

In this case, it is easily shown that there is a conserved quantity, $S+I-\frac{\gamma}{\beta}\ln S+\frac{\lambda}{\beta}\ln I$, which reduces to (\ref{eqn:finalsizepre}) when the vaccination rate is zero (i.e. $\lambda=0$).  Unfortunately, having the conserved quantity is not sufficient to compute the overshoot, since there does not appear to be a way to separate infected and vaccinated individuals when trying to extend the previous calculation. Therefore, we turn to numerical computation (Figure \ref{fig:lambdaS}a). We find that the maximal overshoot is bounded above by the value already obtained in the model without vaccinations. As shown in Figure \ref{fig:lambdaS}, the overshoot has a complicated dependence on the vaccination parameter $\lambda$ and $R_0$.

\noindent \textit{Maximal Overshoot Under Addition of a Risk-Driven Vaccination Rate}

Lastly consider a vaccination rate that is proportional to the number of infected individuals. Such risk-driven behavior may arise for a variety of reasons, including initial vaccine hesitancy, a delay in vaccine availability, or a correlation between willingness to get vaccinated and the number of infected individuals. Consider the following SIRV model:
\begin{align}
\frac{dS}{dt}&=-\beta S I -\lambda S I \label{eqn:SdotSIRV}\\
\frac{dI}{dt}&=\beta S I-\gamma I \label{eqn:IdotSIRV}\\
\frac{dR}{dt}&=\gamma I \\
\frac{dV}{dt} &= \lambda S I \label{eqn:VdotSIRV}
\end{align}
Since the model now has an additional compartment, V, compared with the original SIR model, we must update our definition for overshoot accordingly. Fundamentally, overshoot compares the fraction of people who have not been infected at the epidemic peak and the people who have not been infected at the end of the epidemic. The fraction of people who have not been infected at any particular time, $t$, is $S_t+V_t$. Thus, overshoot can be redefined as follows.
\begin{align}
Overshoot = (S_{t^*} + V_{t^*}) - (S_{\infty} + V_{\infty}) \nonumber
\end{align}
Since the equation for $\frac{dI}{dt}$ remains unchanged, $S_{t^*}=\frac{1}{R_0}$ still applies. Thus, the overshoot equation for models with vaccinated compartments is given by: 
\begin{align}
Overshoot = (\frac{1}{R_0} + V_{t^*}) - (S_{\infty} + V_{\infty}) \label{eqn:overshoot_withvac}
\end{align}
To maximize overshoot, we thus need to find expressions for $R_0, V_{t^*}$, and  $V_{\infty}$ in terms of $S_{\infty}$.

To find $R_0$ we start by taking the ratio $\frac{dI}{dS}$ and integrating as before. It follows that $I + \frac{\beta}{\beta+\lambda}S - \frac{\gamma}{\beta+\lambda} \ln  S$ is constant along any trajectory. Considering the beginning and the end of the epidemic yields:
\begin{align}
I_0 + \frac{\beta}{\beta+\lambda}S_0 - \frac{\gamma}{\beta+\lambda} \ln  S_0 &= I_{\infty} + \frac{\beta}{\beta+\lambda}S_{\infty} - \frac{\gamma}{\beta+\lambda} \ln  S_{\infty} \nonumber
\end{align} 
Using the same initial conditions, asymptotic behavior, and parameter substitution as before ($S_0=1-\epsilon, I_0=\epsilon, I_{\infty}=0, R_0 = \frac{\beta}{\gamma}$) yields the following final size relation.
\begin{align}
R_0 &= \frac{\ln (S_{\infty})}{S_{\infty}-1} \label{eqn:R0expressionVac}
\end{align} 
Thus, we see that $R_0$ for this SIRV model takes on the same expression as the SIR model (\ref{eqn:R0Expression}).

\indent To find $V_{t^*}$, let us take the ratio of time derivatives of the S and V compartments (\ref{eqn:SdotSIRV}), (\ref{eqn:VdotSIRV}),
\begin{align*}
\frac{\frac{dS}{dt}}{\frac{dV}{dt}} &= \frac{-\beta SI -\lambda SI}{\lambda SI} \\
{\frac{dS}{dV}} &= -(\frac{\beta+\lambda}{\lambda})
\end{align*}
from which it follows on integration that
$S + (\frac{\beta+\lambda}{\lambda}) V$ is constant along any trajectory. Considering the beginning and the peak of the epidemic yields:
%\just label{eqn:constantSV}
\begin{align*}
S_0 + (\frac{\beta+\lambda}{\lambda}) V_0 = S_{t^*} + (\frac{\beta+\lambda}{\lambda}) V_{t^*}
\end{align*}
Using the initial conditions ($S_0=1-\epsilon, I_0 = \epsilon, V_0=0$) and recalling that $S_{t^*}=\frac{1}{R_0}$, we obtain the following formula for $V_{t^*}$.
\begin{align} 
V_{t^*} = (1-\frac{1}{R_0})(\frac{\lambda}{\beta+\lambda}) \label{eqn:Vt_eqn}
\end{align}

\indent To find $V_{\infty}$, recall that $S + (\frac{\beta+\lambda}{\lambda}) V$ is constant along any trajectory. Considering the peak of the epidemic and the end of the epidemic yields
\begin{align*}
S_{t^*} + (\frac{\beta+\lambda}{\lambda}) V_{t^*} = S_{\infty} + (\frac{\beta+\lambda}{\lambda}) V_{\infty}
\end{align*}
Using the equation for $V_{t^*}$ (\ref{eqn:Vt_eqn}) and recalling that $S_{t^*}=\frac{1}{R_0}$, we obtain the following equation for $V_{\infty}$.
\begin{align}
V_{\infty} = (1- S_{\infty}) (\frac{\lambda}{\beta+\lambda}) \label{eqn:Vinf}
\end{align}

Substituting the expressions for $R_0$ (\ref{eqn:R0expressionVac}), $V_{t^*}$ (\ref{eqn:Vt_eqn}), $V_{\infty}$ (\ref{eqn:Vinf}) into the overshoot equation (\ref{eqn:overshoot_withvac}) yields:
\begin{align}
Overshoot &= (\frac{S_{\infty}-1}{\ln (S_{\infty})} - S_{\infty})(1-\frac{\lambda}{\beta+\lambda}) \label{eqn:overshootSV}
\end{align}
We see that this expression for the overshoot is simply the overshoot expression for the original SIR model (\ref{eqn:overshoot_sub}) scaled by a factor $1-\frac{\lambda}{\beta+\lambda}$.

\begin{align}
Overshoot_{SIRV(\lambda SI)} &= Overshoot_{SIR} (1-\frac{\lambda}{\beta+\lambda})
\end{align}
Since both $\beta,\lambda \geq 0$, then the factor $1-\frac{\lambda}{\beta+\lambda}$ can never be greater than 1. This implies that the bound on maximal overshoot given by the theorem holds, becoming exact in the limit of no vaccinations (ie. $\lambda=0$). For this model, the maximal overshoot decreases as a function of $\lambda$ in a nonlinear way and has a nonlinear dependence on $R_0$ (Figure \ref{fig:lambda}). 

\subsection*{The Ratio of Overshoot to Outbreak Size}
In the main text, we consider the calculation of overshoot alone. It is also interesting to ask how the overshoot compares to the final attack rate given by the outbreak size. It turns out we can do the calculation analytically using the previous definition for $Overshoot= \frac{1}{R_0} -S_{\infty}$ (\ref{eqn:overshoot}) and defining the total outbreak size as $\textit{Outbreak Size}= 1-S_{\infty}=R_{\infty}$.

Taking the ratio of the two definitions yields:
\begin{equation}
\frac{Overshoot}{\textit{Outbreak Size}}=\frac{\frac{1}{R_0}-S_{\infty}}{1-S_{\infty}} =\frac{1}{R_0(1-S_{\infty})}-\frac{S_{\infty}}{1-S_{\infty}}   
\end{equation}

Substituting $R_0$ using the relationship given by (\ref{eqn:R0Expression}) yields:
\begin{equation}
\frac{Overshoot}{\textit{Outbreak Size}}=\frac{-1}{\ln S_{\infty}}-\frac{S_{\infty}}{1-S_{\infty}} 
\end{equation}

Differentiating this equation with respect to $S_{\infty}$ and setting it to zero to find the extremal points $S_{\infty}^*$ yields:
\begin{equation}
\frac{d(\frac{Overshoot}{\textit{Outbreak Size}})}{dS_{\infty}} = 0 = \frac{1}{S_{\infty}^* (\ln S_{\infty}^* )^2}-\frac{1}{(1-S_{\infty}^*)^2}
\end{equation}

It can be seen upon inspection that the only real solution for $(1-S_{\infty}^* )^2=S_{\infty}^* (\ln S_{\infty}^* )^2$ is at the point $S_{\infty}^*=1.$ This only occurs in the limit of $R_0=1$. Thus, at $R_0$ = 1, the overshoot exactly equals the outbreak size. Then, the overshoot becomes a strictly decreasing fraction of the total outbreak size with increasing $R_0$.

It can be shown that the only real solution to $(1-S_{\infty}^* )^2=S_{\infty}^* (\ln S_{\infty}^* )^2$ for $0\leq S_{\infty}^*\leq 1$ is at the point $S_{\infty}^*=1.$

Since $S_{\infty}^*=0$ is clearly not a solution, we rule that out. Since $(1-S_{\infty}^* )^2=S_{\infty}^* (\ln S_{\infty}^* )^2$ at $S_{\infty}^*=1$, it suffices to show that $f(S_{\infty}^*)\colon= (1-S_{\infty}^* )^2- S_{\infty}^* (\ln S_{\infty}^* )^2> 0$ for all $0< S_{\infty}^*<1$. Since $f'(S_{\infty}^*=1)=0$,  it suffices to show that $f''(S_{\infty}^*)=\frac{2(S_{\infty}^*-1-\ln  S_{\infty}^*)}{S_{\infty}^*}>0$ for all $0< S_{\infty}^*<1$. Since $\ln x<x-1$ for all $x\neq 1$, then it follows that the second derivative must be positive.

The solution $S_{\infty}^*=1$ only occurs in the limit of $R_0=1$. Thus, at $R_0=1$, the overshoot exactly equals the outbreak size. Then, the overshoot becomes a strictly decreasing fraction of the total outbreak size with increasing $R_0$.

While the overshoot is a non-monotonic function of $R_0$, in contrast, the ratio of overshoot to outbreak size is a strictly decreasing function of $R_0$.

% \section*{Supplementary Information}
\setcounter{figure}{0}
\renewcommand{\figurename}{Figure}
\renewcommand{\thefigure}{A\arabic{figure}}

% \newpage
\subsection*{Supplemental Figures}

\subsection*{Figure \ref{fig:Manaus_Rt}. Effective reproduction number ($R_t$) in Manaus, Brazil in 2020 as a function of Date of Symptom Onset.}
\begin{figure}[ht]
\centering
\includegraphics[scale=0.3]{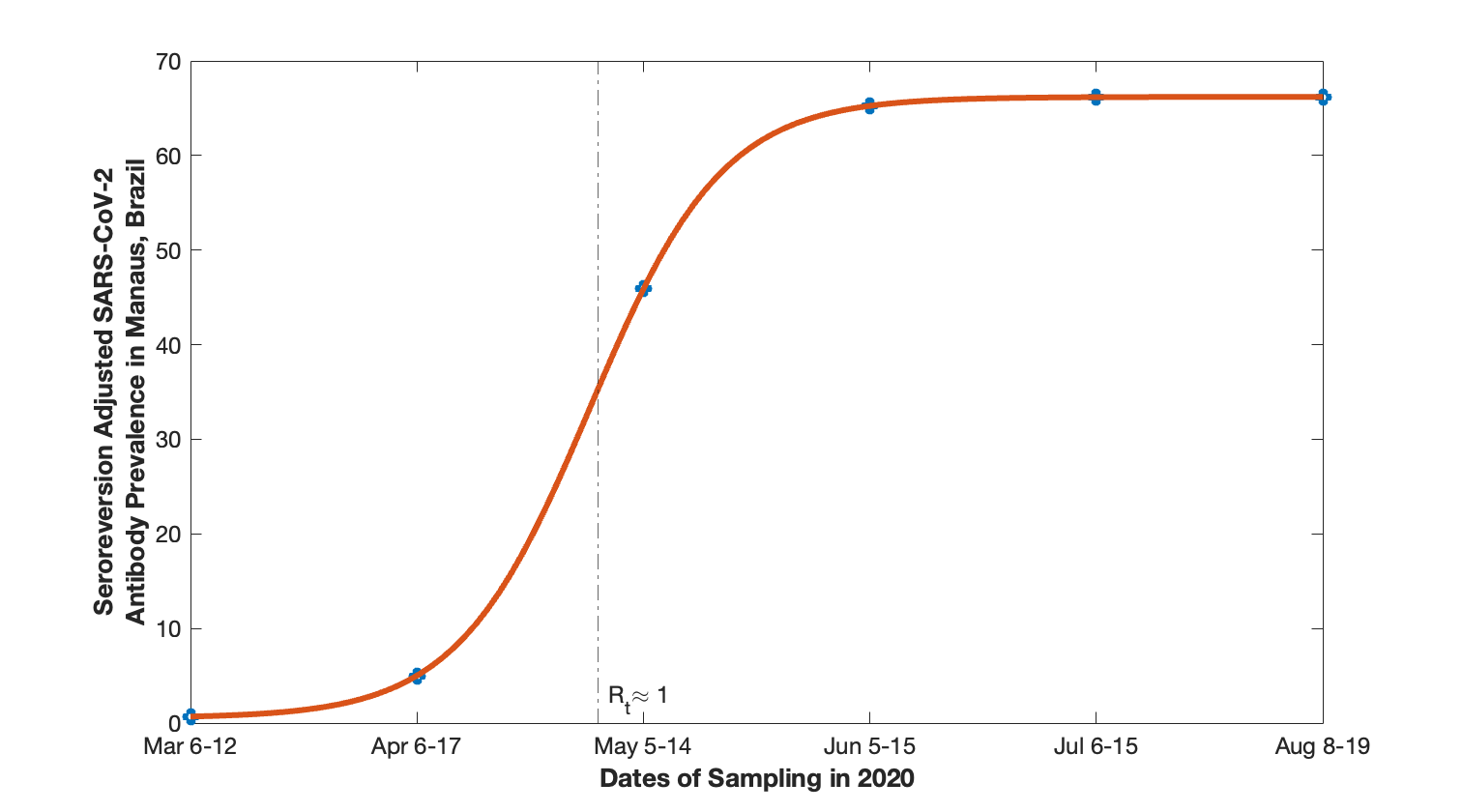}
\caption{Mean cumulative antibody prevalence in Manaus, Brazil in 2020. Seroreversion adjustment done with a 1.4 S/C threshold. Figure adapted from the red points in Figure 2A and Table S2 in Buss et al., “Three-quarters attack rate of SARS-CoV-2 in the Brazilian Amazon during a largely unmitigated epidemic” (\cite{buss_three-quarters_2021}).} \label{fig:seroprev}
\end{figure}

\subsection*{Figure \ref{fig:seroprev}. Cumulative antibody prevalence in Manaus, Brazil in 2020.}
\begin{figure}[ht]
\centering
\includegraphics[scale=0.4]{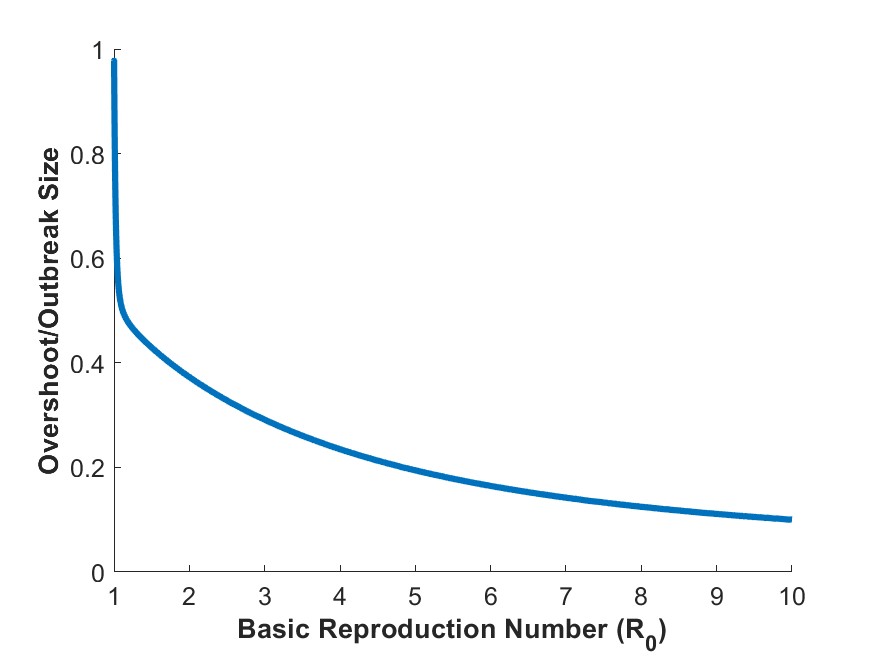}
\caption{The ratio of individuals infected in the overshoot phase compared to total outbreak size as a function of $R_0$.} \label{fig:over_R0}
\end{figure}

\subsection*{Figure \ref{fig:over_R0}. The ratio of individuals infected in the overshoot phase compared to total outbreak size as a function of $R_0$.}
\begin{figure}
\centering
\includegraphics[scale=0.23]{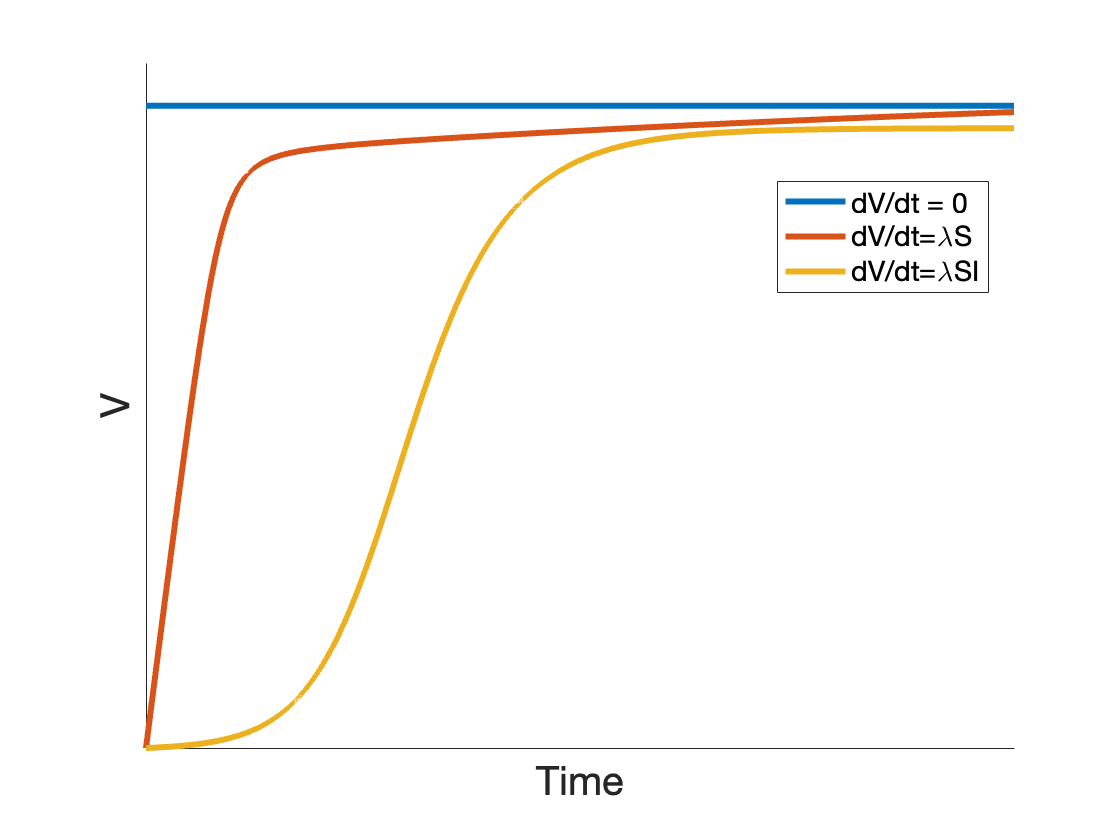}
\caption{The fraction of population that is vaccinated (V) based on different vaccination rates: a vaccination rate of zero over the course of the epidemic (blue), a constant per-capita vaccination rate (red), and a risk-driven vaccination rate (yellow).} \label{fig:dVdt}
\end{figure}

\subsection*{Figure \ref{fig:dVdt}. The fraction of population that is vaccinated (V) based on different vaccination rates.}
\subsection*{Figure \ref{fig:lambdaS}. The overshoot for the SIRV model with $\frac{dV}{dt}=\lambda S$.}
\begin{figure}
\centering
\includegraphics[scale=0.3]{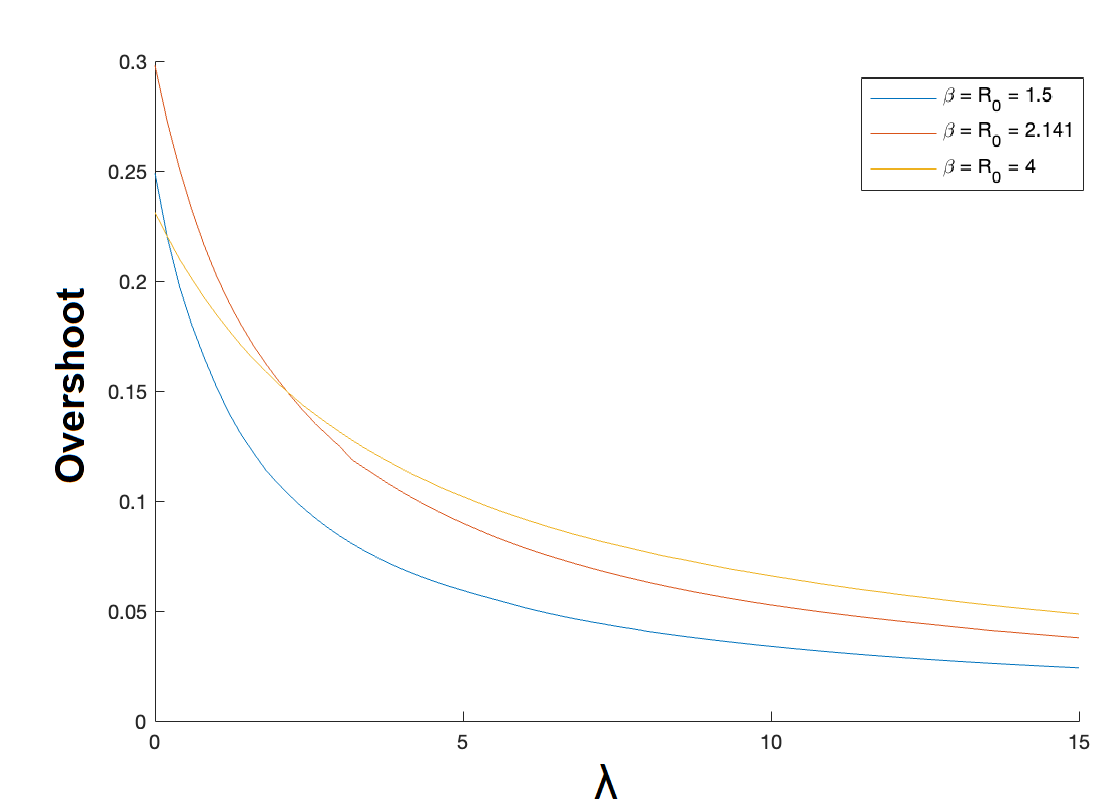}
\caption{The overshoot for the SIRV model with $\frac{dV}{dt}=\lambda SI$ as a function of $\lambda$ for different levels of $\beta$ (or equivalently $R_0$).} \label{fig:lambda}
\end{figure}

\subsection*{Figure \ref{fig:lambda}. The overshoot for the SIRV model with $\frac{dV}{dt}=\lambda SI$.}

\begin{figure}[ht]
\centering
\includegraphics[scale=0.3]{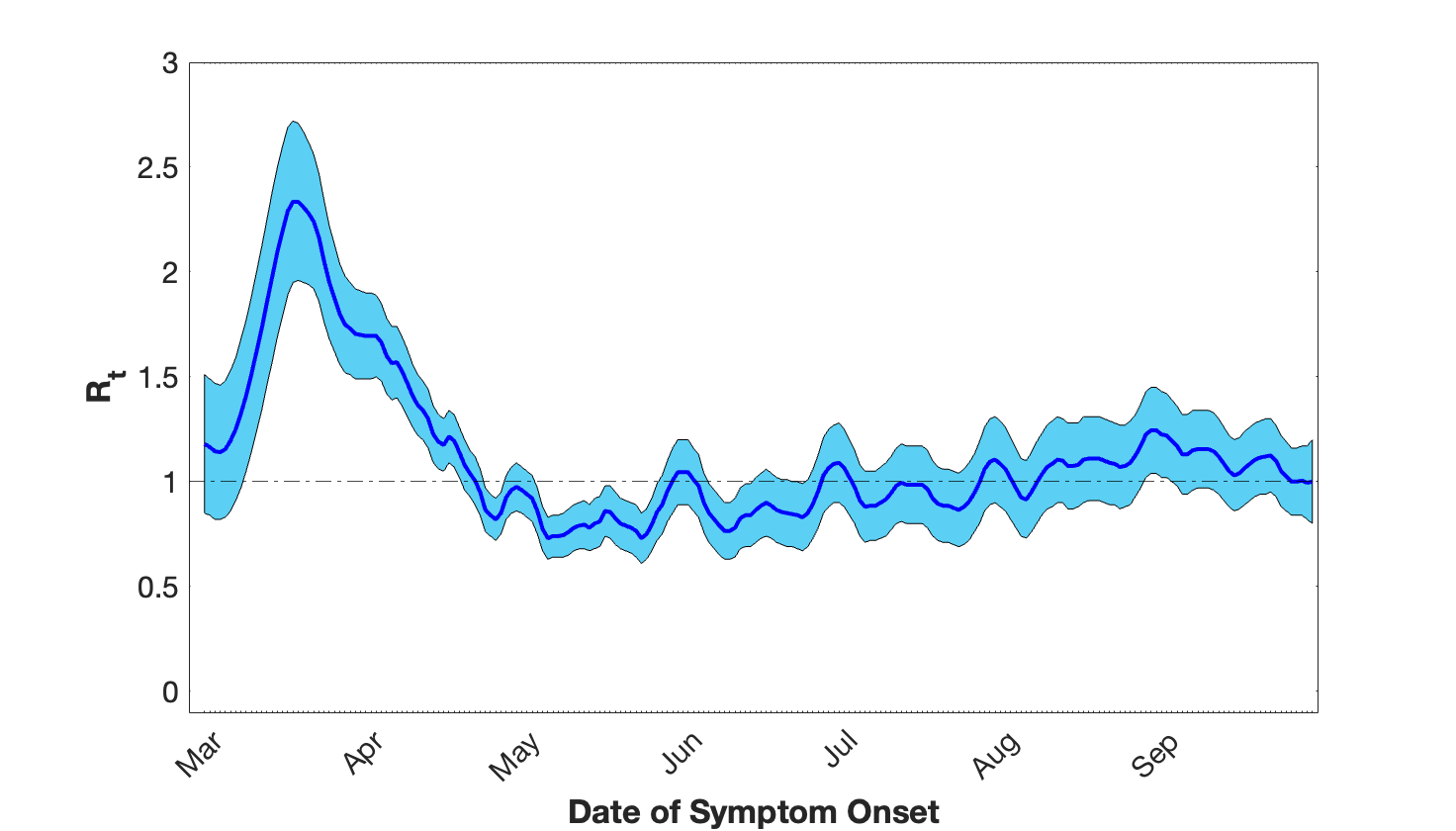}
\caption{Effective reproduction number ($R_t$) in Manaus, Brazil in 2020 as a function of Date of Symptom Onset. Light blue indicates $95\%$ confidence interval around dark blue mean. Figure adapted from Figure S7.D in Buss et al., “Three-quarters attack rate of SARS-CoV-2 in the Brazilian Amazon during a largely unmitigated epidemic” (\cite{buss_three-quarters_2021}).} \label{fig:Manaus_Rt}
\end{figure}

\begin{figure}[ht]
\centering
\includegraphics[scale=0.27]{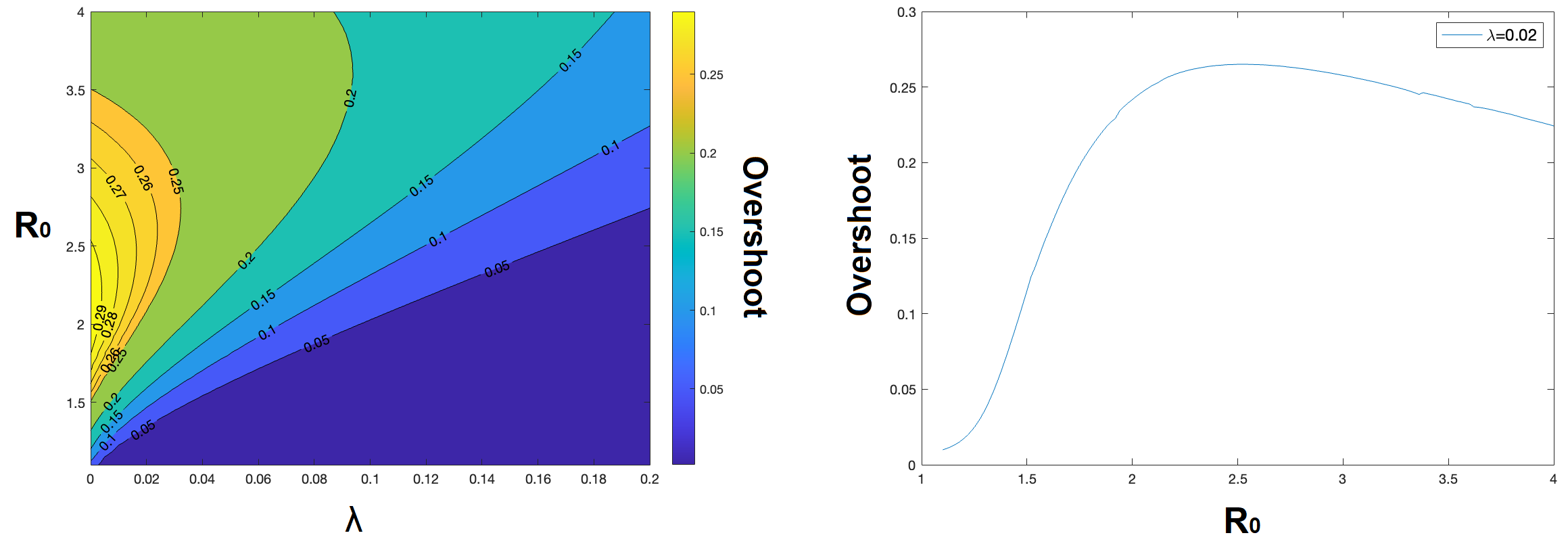}
\caption{a) Contour plot for the overshoot for the SIRV model with $\frac{dV}{dt}=\lambda S$  as a function of $\lambda$ and $R_0$. b) Vertical cross-section of the contour plot from (a) for $\lambda=0.02$.} \label{fig:lambdaS}
\end{figure}

\clearpage
\subsection*{Code to Generate Figures}
Code executed in MATLAB R2022b
\UseRawInputEncoding

\begin{lstlisting}[
frame=single,
numbers=left,
style=Matlab-Pyglike]
%%%%%%%%%%%%%%%%%%%%%%%%%%%%%%%%%%%%%%%%%%%%%%%
% Figure 1
tspan = 0:0.001:50;
I0_frac = 0.001;
vFrac = 0;
R_final = zeros(length(vFrac),1);
OvershootCalc = zeros(length(20),1);
for x = 1:length(vFrac)
    v_frac = vFrac(x);
    SIR_0 = [1-I0_frac-v_frac; I0_frac; 0; vFrac];
    for y = 1%:20
        beta = 1.5;%.*(1-vFrac); %y; %2.15128 ./ (1-v_frac);
        gamma = 1;
        [t,SIR] = ode45(@(t,SIR) SIR_dynamics(t,SIR,beta,gamma,v_frac), tspan, SIR_0);
        HIT_time = find(SIR(:,2) == max(SIR(:,2)));
        S_star = SIR(HIT_time,1);
        S_inf = SIR(length(SIR),1);
        R_final(x,1) = SIR(length(SIR),3);
        OvershootCalc(y,1) = S_star - S_inf;
    end
    disp(x)
end

%%%%%%%%%%%%% Plot 1a: Overshoot in terms of S
close all
figure
subplot(1,2,1)
hold on
plot(t(:,1),SIR(:,1),'b-');
plot(t(:,1),SIR(:,2),'r-');
plot(t(:,1),SIR(:,3),'g-');
ylim([0 1])
set(gca,'XTick',[])
%% Plot special lines
xl = xline(t(HIT_time),'-.',{'t^*'},'FontSize', 15);
xl.LabelVerticalAlignment = 'top';
xl.LabelOrientation = 'horizontal';
yline(S_star,'-.',{'S(t^*)'},'FontSize', 15)
yline(S_inf,'-.',{'S(t=\infty)'},'FontSize', 15)
%% Labels
xlabel('Time','FontSize', 30,'FontWeight','bold')
set(gca,'TickLabelInterpreter', 'tex','FontSize', 15)
ylabel('Fraction of Population','FontSize', 30,'FontWeight','bold')
set(gca,'TickLabelInterpreter', 'tex','FontSize', 15)
legend('S','I','R','FontSize', 15)%,'R')

%%%%%%%%%%%%%%%%% Plot 1b: Overshoot in terms of incidence
%figure
subplot(1,2,2)
hold on
set(gca,'XTick',[])
set(gca,'TickLabelInterpreter', 'tex','FontSize', 15)
yyaxis left
ylabel('Fraction of Population','FontSize', 30,'FontWeight','bold')
set(gca,'TickLabelInterpreter', 'tex','FontSize', 15)
plot(t(:,1),SIR(:,1),'b-');
plot(t(:,1),SIR(:,2),'r-');
plot(t(:,1),beta.*SIR(:,1).*SIR(:,2),'r-.');
ylim([0 1])
%% Plot special lines
xl = xline(t(HIT_time),'-.',{'t^*'},'FontSize', 15);
xl.LabelVerticalAlignment = 'top';
xl.LabelOrientation = 'horizontal';
shaded = area(t(HIT_time:end,1), beta.*SIR(HIT_time:end,1).*SIR(HIT_time:end,2), 'FaceColor', [1, 1, 0]);
%% Labels
xlabel('Time','FontSize', 20,'FontWeight','bold')
legend('S','I','\betaSI','FontSize', 15)%,'R')
yyaxis right
ylabel('\betaSI', 'FontSize',20,'FontWeight','bold')



%%%%%%%%%%%%%%%%%%%%%%%%%%%%%%%%%%%%%%%%%%%%%%%
% Figure 2
%%%%%%%%%%%%%%%%%%%%%%%%%%%%%%%%%%%%%%%%%%%%%%%%%%
%%%%%%%%%%%%%%%%%% Overshoot as function of R_0

tspan = 0:0.001:50;
I0_frac = 0.001;
SIR_0 = [1-I0_frac; I0_frac; 0; I0_frac];
beta_range = 1.05:0.01:10;
gamma_range = [1];
vFrac = 0;

overshoot_vec = zeros(length(beta_range),length(gamma_range));
R_0_vec = zeros(length(beta_range),length(gamma_range));
overshoot2FAR = zeros(length(beta_range),length(gamma_range));
FAR = zeros(length(beta_range),length(gamma_range));
for y = 1:length(gamma_range)
    for x = 1:length(beta_range)
        [t,SIR] = ode45(@(t,SIR) SIR_dynamics(t,SIR,beta_range(1,x),gamma_range(1,y),vFrac), tspan, SIR_0);

        HIT_time = find(SIR(:,2) == max(SIR(:,2)));
        S_star = SIR(HIT_time,1);
        S_inf = SIR(length(SIR),1);
        overshoot_vec(x,y) = S_star - S_inf;
        R_0_vec(x,y) = beta_range(1,x) ./ gamma_range(1,y);
        overshoot2FAR(x,y) = (S_star - S_inf)./(1-S_inf);
        FAR(x,y) = 1-S_inf;
    end
end

figure
plot(R_0_vec(:,y),overshoot_vec(:,y),'-')
ylim([0 0.35])
hold on
max_over = find(overshoot_vec(:,y) == max(overshoot_vec(:,y)));
xl = xline(R_0_vec(max_over),'r-.',{'R_0^*'},'FontSize', 12);
xl.LabelVerticalAlignment = 'bottom';
xl.LabelOrientation = 'horizontal';
yl = yline(max(overshoot_vec(:,y)),'r-.',{'Overshoot^*'},'FontSize', 12);
xlabel('R_0','FontSize', 15)
ylabel('Overshoot','FontSize', 15)


%%%%%%%%%%%%%%%%%%%%%%%%%%%%%%%%%%%%%%%%%%%%%%%
% Figure A1

% Data given by Direct Correspondence with authors of Buss et. al, Science 2020, “Three-quarters attack rate of SARS-CoV-2 in the Brazilian Amazon during a largely unmitigated epidemic”. 

% This figure is an adaptation of S7.D in their Supplementary Material

% Raw data compiled from Rt_Science_buss -> symptoms_csv -> tdates_MN_symp.csv & Rlow_MN_symp.csv & Rhigh_MN_symp.csv into following Table.



Time Point	Lower Bound on 95% CI for Rt	Upper Bound on 95% CI for Rt	Date
1	1.35	3.08	2/3/20
2	1.38	3.02	2/4/20
3	1.39	2.92	2/5/20
4	1.37	2.8	2/6/20
5	1.34	2.67	2/7/20
6	1.3	2.54	2/8/20
7	1.26	2.42	2/9/20
8	1.21	2.3	2/10/20
9	1.18	2.2	2/11/20
10	1.15	2.12	2/12/20
11	1.12	2.05	2/13/20
12	1.1	1.99	2/14/20
13	1.07	1.93	2/15/20
14	1.05	1.89	2/16/20
15	1.03	1.84	2/17/20
16	1.01	1.8	2/18/20
17	0.98	1.75	2/19/20
18	0.96	1.72	2/20/20
19	0.95	1.68	2/21/20
20	0.93	1.65	2/22/20
21	0.92	1.63	2/23/20
22	0.91	1.62	2/24/20
23	0.91	1.6	2/25/20
24	0.9	1.59	2/26/20
25	0.89	1.57	2/27/20
26	0.88	1.55	2/28/20
27	0.87	1.53	2/29/20
28	0.85	1.51	3/1/20
29	0.84	1.49	3/2/20
30	0.82	1.47	3/3/20
31	0.82	1.46	3/4/20
32	0.83	1.48	3/5/20
33	0.86	1.53	3/6/20
34	0.91	1.59	3/7/20
35	0.97	1.68	3/8/20
36	1.05	1.77	3/9/20
37	1.14	1.88	3/10/20
38	1.24	2	3/11/20
39	1.34	2.12	3/12/20
40	1.46	2.25	3/13/20
41	1.57	2.38	3/14/20
42	1.69	2.5	3/15/20
43	1.79	2.6	3/16/20
44	1.89	2.69	3/17/20
45	1.95	2.72	3/18/20
46	1.96	2.71	3/19/20
47	1.95	2.67	3/20/20
48	1.94	2.62	3/21/20
49	1.92	2.56	3/22/20
50	1.86	2.47	3/23/20
51	1.76	2.34	3/24/20
52	1.68	2.22	3/25/20
53	1.62	2.13	3/26/20
54	1.56	2.04	3/27/20
55	1.52	1.98	3/28/20
56	1.51	1.95	3/29/20
57	1.49	1.92	3/30/20
58	1.49	1.91	3/31/20
59	1.49	1.9	4/1/20
60	1.49	1.9	4/2/20
61	1.5	1.89	4/3/20
62	1.48	1.85	4/4/20
63	1.42	1.78	4/5/20
64	1.39	1.74	4/6/20
65	1.4	1.74	4/7/20
66	1.36	1.69	4/8/20
67	1.31	1.63	4/9/20
68	1.26	1.56	4/10/20
69	1.22	1.51	4/11/20
70	1.2	1.48	4/12/20
71	1.16	1.44	4/13/20
72	1.09	1.36	4/14/20
73	1.06	1.32	4/15/20
74	1.05	1.3	4/16/20
75	1.09	1.34	4/17/20
76	1.07	1.32	4/18/20
77	1.02	1.26	4/19/20
78	0.96	1.2	4/20/20
79	0.93	1.15	4/21/20
80	0.89	1.12	4/22/20
81	0.84	1.06	4/23/20
82	0.76	0.97	4/24/20
83	0.74	0.94	4/25/20
84	0.72	0.92	4/26/20
85	0.75	0.95	4/27/20
86	0.82	1.03	4/28/20
87	0.85	1.07	4/29/20
88	0.86	1.09	4/30/20
89	0.85	1.07	5/1/20
90	0.83	1.05	5/2/20
91	0.81	1.03	5/3/20
92	0.75	0.97	5/4/20
93	0.67	0.88	5/5/20
94	0.63	0.83	5/6/20
95	0.64	0.84	5/7/20
96	0.64	0.84	5/8/20
97	0.64	0.85	5/9/20
98	0.65	0.87	5/10/20
99	0.67	0.89	5/11/20
100	0.68	0.9	5/12/20
101	0.68	0.91	5/13/20
102	0.67	0.89	5/14/20
103	0.68	0.92	5/15/20
104	0.69	0.93	5/16/20
105	0.74	0.98	5/17/20
106	0.73	0.98	5/18/20
107	0.7	0.95	5/19/20
108	0.68	0.92	5/20/20
109	0.67	0.91	5/21/20
110	0.66	0.9	5/22/20
111	0.64	0.89	5/23/20
112	0.61	0.85	5/24/20
113	0.63	0.87	5/25/20
114	0.67	0.92	5/26/20
115	0.72	0.99	5/27/20
116	0.75	1.03	5/28/20
117	0.81	1.1	5/29/20
118	0.85	1.16	5/30/20
119	0.89	1.2	5/31/20
120	0.89	1.2	6/1/20
121	0.89	1.2	6/2/20
122	0.86	1.16	6/3/20
123	0.83	1.13	6/4/20
124	0.76	1.04	6/5/20
125	0.71	0.99	6/6/20
126	0.68	0.96	6/7/20
127	0.65	0.93	6/8/20
128	0.63	0.9	6/9/20
129	0.63	0.9	6/10/20
130	0.64	0.92	6/11/20
131	0.68	0.97	6/12/20
132	0.69	0.99	6/13/20
133	0.69	0.99	6/14/20
134	0.71	1.02	6/15/20
135	0.73	1.04	6/16/20
136	0.74	1.06	6/17/20
137	0.73	1.04	6/18/20
138	0.71	1.02	6/19/20
139	0.7	1.01	6/20/20
140	0.69	1.01	6/21/20
141	0.69	1	6/22/20
142	0.68	1	6/23/20
143	0.67	0.99	6/24/20
144	0.69	1.01	6/25/20
145	0.73	1.06	6/26/20
146	0.78	1.13	6/27/20
147	0.85	1.21	6/28/20
148	0.88	1.25	6/29/20
149	0.9	1.27	6/30/20
150	0.9	1.28	7/1/20
151	0.88	1.25	7/2/20
152	0.84	1.2	7/3/20
153	0.8	1.15	7/4/20
154	0.74	1.08	7/5/20
155	0.71	1.05	7/6/20
156	0.72	1.05	7/7/20
157	0.72	1.05	7/8/20
158	0.73	1.07	7/9/20
159	0.74	1.09	7/10/20
160	0.77	1.12	7/11/20
161	0.8	1.16	7/12/20
162	0.81	1.18	7/13/20
163	0.8	1.17	7/14/20
164	0.8	1.17	7/15/20
165	0.8	1.17	7/16/20
166	0.8	1.17	7/17/20
167	0.78	1.15	7/18/20
168	0.74	1.1	7/19/20
169	0.72	1.07	7/20/20
170	0.71	1.06	7/21/20
171	0.71	1.06	7/22/20
172	0.7	1.05	7/23/20
173	0.69	1.04	7/24/20
174	0.7	1.06	7/25/20
175	0.72	1.09	7/26/20
176	0.76	1.13	7/27/20
177	0.8	1.19	7/28/20
178	0.86	1.26	7/29/20
179	0.89	1.3	7/30/20
180	0.9	1.31	7/31/20
181	0.88	1.29	8/1/20
182	0.86	1.26	8/2/20
183	0.82	1.21	8/3/20
184	0.78	1.16	8/4/20
185	0.74	1.11	8/5/20
186	0.73	1.1	8/6/20
187	0.76	1.14	8/7/20
188	0.8	1.19	8/8/20
189	0.84	1.23	8/9/20
190	0.87	1.27	8/10/20
191	0.88	1.29	8/11/20
192	0.9	1.31	8/12/20
193	0.9	1.3	8/13/20
194	0.87	1.28	8/14/20
195	0.87	1.28	8/15/20
196	0.88	1.28	8/16/20
197	0.9	1.31	8/17/20
198	0.91	1.31	8/18/20
199	0.91	1.31	8/19/20
200	0.91	1.31	8/20/20
201	0.9	1.3	8/21/20
202	0.89	1.29	8/22/20
203	0.89	1.28	8/23/20
204	0.87	1.27	8/24/20
205	0.88	1.27	8/25/20
206	0.89	1.29	8/26/20
207	0.93	1.32	8/27/20
208	0.97	1.37	8/28/20
209	1.02	1.43	8/29/20
210	1.04	1.45	8/30/20
211	1.04	1.45	8/31/20
212	1.02	1.43	9/1/20
213	1.02	1.42	9/2/20
214	1	1.39	9/3/20
215	0.98	1.36	9/4/20
216	0.94	1.32	9/5/20
217	0.94	1.32	9/6/20
218	0.96	1.34	9/7/20
219	0.97	1.34	9/8/20
220	0.97	1.34	9/9/20
221	0.97	1.34	9/10/20
222	0.96	1.33	9/11/20
223	0.94	1.3	9/12/20
224	0.91	1.26	9/13/20
225	0.88	1.22	9/14/20
226	0.86	1.2	9/15/20
227	0.87	1.21	9/16/20
228	0.89	1.24	9/17/20
229	0.91	1.26	9/18/20
230	0.93	1.28	9/19/20
231	0.94	1.29	9/20/20
232	0.94	1.3	9/21/20
233	0.95	1.3	9/22/20
234	0.93	1.27	9/23/20
235	0.88	1.22	9/24/20
236	0.86	1.19	9/25/20
237	0.84	1.16	9/26/20
238	0.84	1.16	9/27/20
239	0.84	1.17	9/28/20
240	0.82	1.17	9/29/20
241	0.8	1.2	9/30/20



close all
Buss_RtManaus = table2array(WaveTable(:,1:3))'; %Import table above as “WaveTable”

figure
x_allData = Buss_RtManaus(1,:);
y_allData = mean(Buss_RtManaus([2,3],:));
x = x_allData(1,28:end); %Only plot starting from March 1 (which is data point 28 out of the full data set)
y = y_allData(1,28:end);
curve1 = Buss_RtManaus(2,28:end);
curve2 = Buss_RtManaus(3,28:end);
x2 = [x, fliplr(x)];
inBetween = [curve1, fliplr(curve2)];
fill(x2, inBetween, [0.3569,0.8118,0.9569]);
hold on;
plot(x, y, 'b', 'LineWidth', 2); 
xticks(28:27+length(x)) 
xticklabels(table2cell(RlowMNsymp(28:end,4)));
xtickangle(45)
xlabel('Date of Symptom Onset','FontSize', 24, 'FontWeight','bold')
set(gca,'TickLabelInterpreter', 'tex','FontSize', 15)
ylabel('R_t','FontSize', 24, 'FontWeight','bold')
set(gca,'TickLabelInterpreter', 'tex','FontSize', 15)
ylim([-0.1, 3])
xlim([25, 242])
ax = gca;
ax.TickLength = [0.001,0.001];
yline(1,'k-.','FontSize',12)




%%%%%%%%%%%%%%%%%%%%%%%%%%%%%%%%%%%%%%%%%%%%%%%
% Figure A2
%%%%%%%%%%%%

%% Data taken from table S2 in Supplemental Information of Buss et. al, Science 2020, “Three-quarters attack rate of SARS-CoV-2 in the Brazilian Amazon during a largely unmitigated epidemic”.


Prevalence_seroreversionAdj = [0.7, 5, 45.9, 65.2, 66.2, 66.2];%, 72.2, 76];
Prevalence_seroreversionAdj_lowerbound = [0.2, 3.7, 41.8, 60.5, 61.5, 61.8, 64.3, 66.6];
Prevalence_seroreversionAdj_upperbound = [1.5, 6.6, 50.6, 74.3, 80.1, 80.9, 91.8, 97.9];
Prevalence_dates = {'Mar 6-12', 'Apr 6-17', 'May 5-14', 'Jun 5-15', 'Jul 6-15', 'Aug 8-19'}; %, 'Sep 5-14', 'Oct 10-17'};
% Define the logistic function to fit
logisticFun = @(params, x) params(1) + (params(2) - params(1)) ./ (1 + exp(-params(3) .* (x - params(4))));
% Initial guess for the parameters [a, b, c, d] of the logistic function
initialGuess = [0, 1, 1, 5];
% Fit the logistic function using lsqcurvefit
fitParams = lsqcurvefit(logisticFun, initialGuess, 1:length(Prevalence_seroreversionAdj), Prevalence_seroreversionAdj);
% Generate a finer grid of x values for plotting the fitted curve
xFit = linspace(min(1:length(Prevalence_seroreversionAdj)), max(1:length(Prevalence_seroreversionAdj)), 100);
% Calculate the corresponding y values for the fitted curve
yFit = logisticFun(fitParams, xFit);
% Plot the original data points and the fitted curve
figure;
p = plot(1:length(Prevalence_seroreversionAdj), Prevalence_seroreversionAdj, 'o', xFit, yFit, '-', 'LineWidth',3);
xlabel('Dates of Sampling in 2020','FontSize',12,'FontWeight','bold')
xticks(1:length(Prevalence_dates))
ylabel(['Seroreversion Adjusted SARS-CoV-2', newline, 'Antibody Prevalence in Manaus, Brazil'],'FontSize',12,'FontWeight','bold')
%title('Manaus COVID-19 First Wave')
xticklabels(Prevalence_dates);
set(gca,'TickLabelInterpreter', 'tex','FontSize', 12)
hold on
yl = xline(2.8,'-.','R_t\approx 1','FontSize', 12);
yl.LabelVerticalAlignment = 'bottom';
yl.LabelOrientation = 'horizontal';


%%%%%%%%%%%%%%%%%%%%%%%%%%%%%%%%%%%%%%%%%%%%%%%
% Figure A3

close all
% Epidemic Overshoot as a Function of R0 in MATLAB
% Varying R0 values
R0_values = [1:0.001:1.12,1.13:0.01:10];
% Initialize overshoot values
overshoot_values = zeros(size(R0_values));
overshoot_fracvalues = zeros(size(R0_values));
for i = 1:length(R0_values)
    % Parameters
    R0 = R0_values(i);
    beta = R0 * gamma;
    % Solve the ODE system
    odeSystem = @(t, y) [-beta * y(1) * y(2); % dS/dt
        beta * y(1) * y(2) - gamma * y(2); % dI/dt
        gamma * y(2)]; % dR/dt
    [~, y] = ode89(odeSystem, tspan, initialConditions);
    % Find the peak of infections and its corresponding time index
    [maxInfections, peakIndex] = max(y(:, 2));
    % Calculate the epidemic overshoot (difference in fraction of susceptible)
    overshoot = y(peakIndex, 1) - y(end, 1);
    overshoot_values(i) = overshoot;
    overshoot_fracvalues(i) = overshoot ./ (1-y(end, 1));
end

figure
xlabel('Basic Reproduction Number (R_0)', 'FontSize',14, 'FontWeight','bold');
ylabel('Overshoot/Outbreak Size','FontSize',14, 'FontWeight','bold');
hold on
p = plot(R0_values, overshoot_fracvalues, '-.', 'LineWidth',3);
set(gca,'FontSize',12)


%%%%%%%%%%%%%%%%%%%%%%%%%%%%%%%%%%%%%%%%%%%%%%%
% Figure A4

%% Save SIR_dynamics2 as a separate file.
% function dSIR = SIR_dynamics2(t,SIR,beta,gamma,lambda)
% dSIR = zeros(4,1);
% dSIR(1) = -beta*SIR(1)*SIR(2) - lambda*SIR(1);
% dSIR(2) = beta*SIR(1)*SIR(2) - gamma*SIR(2);
% dSIR(3) = gamma*SIR(2);
% dSIR(4) = lambda*SIR(1);
%%

tspan = 0:0.001:20;
I0_frac = 0.001;
vFrac = 0;
gamma = 1;
beta = 2;
lambda = 0.0;
SIR_0 = [1-I0_frac-vFrac; I0_frac; 0; vFrac]
[t,SIR] = ode45(@(t,SIR) SIR_dynamics2(t,SIR,beta,gamma,lambda), tspan, SIR_0);
figure
hold on
plot(t,SIR(:,4)+0.075,'LineWidth',3)
xlabel('Time', 'FontSize', 20)
ylabel('V', 'FontSize', 20)
set(gca,'XTick',[]);
set(gca,'YTick',[]);
lambda = 0.04;
beta = 5;
[t,SIR] = ode45(@(t,SIR) SIR_dynamics2(t,SIR,beta,gamma,lambda), tspan, SIR_0);
hold on
plot(t,SIR(:,4),'LineWidth',3)
lambda = 0.2;
beta = 2;
[t,SIR] = ode45(@(t,SIR) SIR_dynamics3(t,SIR,beta,gamma,lambda), tspan, SIR_0);
hold on
plot(t,SIR(:,4),'LineWidth',3)
legend( 'dV/dt = 0', 'dV/dt=\lambdaS','dV/dt=\lambdaSI','FontSize',14)


%%%%%%%%%%%%%%%%%%%%%%%%%%%%%%%%%%%%%%%%%%%%%%%
% Figure A5

%close all
tspan = 0:0.01:100;
I0_frac = 0.001;
lambda = [0:0.001:0.2]';
betas = [1.1:0.001:4]';
R_final = zeros(length(lambda),length(betas));
OvershootCalc = zeros(length(lambda),length(betas));
VacFrac = zeros(length(lambda),length(betas));

for x = 1:length(lambda)
    lambdaVal = lambda(x);
    SIR_0 = [1-I0_frac; I0_frac; 0; 0];
    for y = 1:length(betas)
        beta = betas(y); % ./ (1-v_frac);
        gamma = 1;
        [t,SIR] = ode45(@(t,SIR) SIR_dynamics2(t,SIR,beta,gamma,lambdaVal), tspan, SIR_0);
        HIT_time = find(SIR(:,2) == max(SIR(:,2)));
        S_star = SIR(HIT_time,1);
        S_inf = SIR(length(SIR),1);
        R_final(x,y) = SIR(length(SIR),3);
        OvershootCalc(x,y) = (SIR(HIT_time,1)+SIR(HIT_time,4)) - (SIR(length(SIR),1)+SIR(length(SIR),4));
        VacFrac(x,y) = SIR(length(SIR),4);
        eps_time = find(SIR(:,2) == 0.001);
    end
end

%% FIGURE FOR lambda S model
figure
subplot(1,2,1)
[Xs,Ys]=meshgrid(lambda,betas);
%[Xs,Ys]=meshgrid(lambda(1:31),betas);
contourf(Xs',Ys',OvershootCalc,[0,0.05,0.1,0.15,0.2,0.25:0.01:0.3],"ShowText",true)
%contourf(Xs',Ys',OvershootCalc(1:31,:),"ShowText",true)
xlabel('\lambda','FontSize',20,'FontWeight','bold')
ylabel('R_0','FontSize', 20, 'FontWeight', 'bold')
set(gca,'FontSize', 15)
%title('Overshoot')
c = colorbar;
c.Label.String = 'Overshoot';
%% Cross-section of Contour Plot
subplot(1,2,2)
plot(betas./gamma,OvershootCalc(5,:))
xlabel('R_0', 'FontSize', 16, 'FontWeight', 'bold')
ylabel('Overshoot', 'FontSize', 16, 'FontWeight', 'bold')
legend('\lambda=0.02','FontSize',12)




%%%%%%%%%%%%%%%%%%%%%%%%%%%%%%%%%%%%%%%%%%%%%%%
% Figure A6


%%%%%%%%%%%%%%%%%% Overshoot as function of Lambda in dV/dt=lambdaSI vaccine model
tspan = 0:0.001:50;
I0_frac = 0.001;
SIR_0 = [1-I0_frac; I0_frac; 0; 0];
beta_range = [1.5, 2.141, 4];
gamma = 1;
lambda_range = [0:0.1:15];

overshoot_vec = zeros(length(beta_range),length(lambda_range));

for x = 1:length(beta_range)
    for y = 1:length(lambda_range)
        [t,SIR] = ode78(@(t,SIR) SIR_dynamics3(t,SIR,beta_range(x), gamma, lambda_range(y)), tspan, SIR_0);

        HIT_time = find(SIR(:,2) == max(SIR(:,2)));
        S_star = SIR(HIT_time,1)+SIR(HIT_time,4); %Recall overshoot in a vaccine model requires consideration of Susceptibles and vaccinated individuals (eqn A23)
        S_inf = SIR(length(SIR),1)+SIR(length(SIR),4);
        overshoot_vec(x,y) = S_star - S_inf;
    end
end

figure
hold on
plot(1:length(SIR),SIR(:,1))
plot(1:length(SIR),SIR(:,2))
plot(1:length(SIR),SIR(:,3))
plot(1:length(SIR),SIR(:,4))

figure
hold on
plot(lambda_range,overshoot_vec(1,:),'-')
plot(lambda_range,overshoot_vec(2,:),'-')
plot(lambda_range,overshoot_vec(3,:),'-')
xlabel('\lambda','FontSize', 15)
ylabel('Overshoot','FontSize', 15)
legend('\beta=R_0=1.5', '\beta=R_0=2.141', '\beta=R_0=4')
\end{lstlisting}

\end{document}